\documentstyle[12pt]{article}

\newcommand{\sect}[1]{\setcounter{equation}{0}\section{#1}}
\newcommand{\subsect}[1]{\subsection{#1}}

\newfont{\frak}{eufm10 scaled\magstep1}
\newfont{\extra}{msbm10 scaled\magstep1}

\def\be{\begin{equation}}
\def\ee{\end{equation}}
\def\bea{\begin{eqnarray}}
\def\eea{\end{eqnarray}}

\def\b{\beta}
\def\bx{\beta_x}
\def\bt{\beta_t}
\def\bsx{\beta^s_x}
\def\bst{\beta^s_t}

\def\sx{\sigma_x}
\def\st{\sigma_t}

\begin{document}
%%%%%%%%%%%%%%%%%%%%%%%%%%%%%%%%%%%%%%

\begin{center} 
{\LARGE{\bf{Lie symmetries \\[0.45cm]  of difference
equations}}}\footnote{Talk delivered by J. Negro at the DI-CRM Workshop held in
Prague, 18-21.6.2000}
\end{center}

\bigskip\bigskip

\begin{center} 
D. Levi$^{*}$, J. Negro$^{**}$ and M.A. del Olmo$^{**}$ 
\end{center}

\begin{center} 
$^{*}${\sl Departimento di Fisica, Universit\'a  Roma Tre and INFN--Sezione di 
Roma Tre \\ 
Via della Vasca Navale 84, 00146 Roma, Italy}\\ 
\medskip

$^{**}${\sl Departamento de F\'{\i}sica Te\'orica, Universidad de Valladolid,
\\ E-47011, Valladolid, Spain.}\\ 
\medskip

{e-mail:levi@amaldi.fis.uniroma3.it,  jnegro@fta.uva.es, olmo@fta.uva.es} 
\end{center}

\vskip 0.5cm

\bigskip

\begin{abstract} 
The discrete heat equation is worked out in order to illustrate the 
search of symmetries of difference equations. It is paid an special attention
to the Lie structure of these symmetries, as well as to their dependence on the
derivative discretization. The case of $q$--symmetries for discrete equations 
in a $q$--lattice is also considered.
\end{abstract}
\vskip 1cm

%%%%%%%%%%%%%%%%%%%%% INTRODUCTION %%%%%%%%%%%%%%%%%%%%%%%%%%%%%%%%%
\sect{Introduction\label{introduccion}}
%%%%%%%%%%%%%
As it is well known Lie point symmetries were
introduced by  Lie for solving differential equations, 
providing one of the most efficient methods for obtaining exact analytical
solutions of partial differential equations \cite{olver}. The interest for
discrete systems in the last years has led to extend the Lie method
 to the case of discrete equations \cite{levi}--\cite{quispel}.

A general difference equation, involving one scalar function 
$u(x)$ of $p$ independent variables
$x=(x_1,x_2,\dots,x_p)$ evaluated at a finite number of points on a lattice
will be written in the form
\be\label{ecuaciondiscreta}
E(x,\ T^{a}u(x),\ T^{b_i}\Delta_{x_i} u(x),\ T^{c_{ij}}\Delta_{x_i}\Delta_{x_j} 
u(x), \dots)=0 ,
\ee
where  the shift operators $T^{a},T^{b_i},\ T^{c_{ij}}$ are defined by
$$
T^{a}u(x) :=\left\{ T^{a_1}_{x_1}T^{a_2}_{x_2}\cdots T^{a_p}_{x_p}\,
u(x)\right\}_{a_i=m_i}^{n_i},  a=(a_1,a_2,\dots, a_p), 
\quad i=1,2,\dots, p,
$$
with $a_i,\ m_i, \ n_i,\ (m_i \leq n_i)$, fixed integers, 
$$
T^{a_i}_{x_i}u(x)= u(x_1,x_2,\dots,x_{i-1}, x_i+a_i\sigma_i,
x_{i+1},\dots,x_p),
$$
and
$\sigma_i$ is the positive lattice  spacing in the uniform
lattice of the variable $x_i \ (i=1,\dots, p)$.
The other shift operators $T^{b_i},\ T^{c_{ij}}$ are defined in a similar way.
The difference operators $\Delta_{x_i}$ are defined so that in the continuous
limit turn into partial derivatives.  

%%%%%%%%%%%%%%

In the following we will make use of the approach presented in
\cite{decio}, based on the formalism of continuous evolutionary vector fields
\cite{olver}. 
The infinitesimal symmetry vectors  in evolutionary
form for the difference equation of order $N$   given in 
(\ref{ecuaciondiscreta})  take the general expression
\be \label{vectorgeneral}
X_e\equiv Q\partial u= \left( \sum_i \xi _i (x, T^{a}u, \sx,\st) T^{b}
\Delta_{x_i} u -\phi(x,T^{c}u, \sx,\st)
\right)\partial u ,
\ee
where  $\xi _i (x, T^{a}u, \sx,\st)$ and $\phi(x,T^{c}u, \sx,\st)$ are
operator valued functions  which in the continuous limit become the functions 
$\xi _i (x,u)$ and $\phi(x, u)$, respectively, giving rise to
Lie point symmetries.

The vector fields $X_e$ generate the symmetry algebra of the discrete equation 
(\ref{ecuaciondiscreta}), whose elements transform solutions $u(x)$ of the
equation  into solutions $\tilde u(x)$.
The $N$--th prolongation of $X_e$ must verify the invariance condition
\be\label{prolongacion}
pr^N X_e E|_{E=0}=0.
\ee
The prolongation $pr^N X_e$ is 
\be\label{prolongationvector}
pr^N X_e=\sum_a T^a Q\partial_{T^a u} +
\sum_{b_i} T^{b_i}Q^{x_i}\partial_{T^{b_i}\Delta_{x_i}u}  +
\sum_{c_{ij}} T^{c_{ij}}Q^{x_{i}x_{j}}\partial_{T^{c_{ij}}
\Delta_{x_i}\Delta_{x_j}u}  +\dots
\ee
where summations in (\ref{prolongationvector}) are over all the sites present
in (\ref{ecuaciondiscreta}). The symbols $Q^{x_i},\, Q^{{x_i}{x_j}},\dots$ are
total variations of $Q$, i.e.,
$Q^{x_i}=\Delta_{x_i}^T Q, \  
Q^{{x_i}{x_j}}=\Delta_{x_i}^T \Delta_{x_j}^T Q ,\  \cdots,\ $ defined by
$$
\Delta_{x}^T f(x, u(x), \Delta_{x}u(x), \dots)=\frac 1\sigma[f(x+\sigma, 
u(x+\sigma),
(\Delta_{x}u)(x+\sigma), \dots) 
-f(x, u(x), \Delta_{x}u(x), \dots)], 
$$
while the partial variation $\Delta_{x_i}$ is
\be\label{dd}
\Delta_{x} f(x, u(x), \Delta_{x}u(x), {\dots}){=}\frac 1\sigma[f(x{+}\sigma,
u(x), (\Delta_{x}u)(x), \dots) {-}f(x, u(x), \Delta_{x}u(x), {\dots})] .
\ee

The  solutions of (\ref{prolongacion}) gives the symmetries of
equation (\ref{ecuaciondiscreta}) when using the  difference operator
(\ref{dd}). The determining equations for
$\xi _i$  and $\phi$  are obtained by considering linearly independent the
expressions in the discrete derivatives $T^{a}\Delta_{x_i}u$,
$T^{b}\Delta_{x_ix_j}u$, \dots.
The Lie commutators of the vector fields $X_e$ are obtained by commuting their 
first prolongations and projecting onto the symmetry algebra $\cal G$. 

Since we will restrict here to linear equations we can assume that the 
evolutionary vectors (\ref{vectorgeneral}) have the form 
$X_e=  (\hat X u) \partial_u$,
where
$$
\hat X=  \sum_i \xi _i (x, T^{a}, \sx,\st) 
\Delta_{x_i}  -\phi(x,T^{a}, \sx,\st)  .
$$
The operators $\hat X$, in general, may span only a subalgebra of the whole Lie
symmetry algebra \cite{decio}.

%%%%%%%%%%%%%%
In Ref.\ \cite{javier} the symmetries were obtained using the
above mentioned difference operator
\be\label{derivadadiscretap}
\Delta_{x_i}\equiv \Delta^+_{x_i}=\frac{T_{x_i} -1}{\sigma_i},
\ee
which, when $\sigma_i \to 0$, goes into the standard right derivative with
respect to $x_i$. Since other definitions of the difference operator can be
introduced \cite{milne}, one would like to show that the
algebraic structure of the symmetries is independent on the choice undertaken.
In this work we will also use the left derivative
\be\label{derivadadiscretam}
\Delta^-_{x_i}=\frac{1-T_{x_i}^{-1}}{\sigma_i},
\ee
and the symmetric derivative 
(which goes into the derivative with respect to $x_i$ up to terms of order
$\sigma_i^2$)
\be\label{derivadadiscretas}
\Delta^s_{x_i}=\frac{T_{x_i}-T_{x_i}^{-1}}{2\sigma_i}.
\ee

In the following we will see that by an appropriate definition of the Leibniz 
rule we can construct Lie symmetries, in principle, for any difference operator.
In Section \ref{calor} we will introduce this procedure on the example of the
discrete heat equation. We shall study separately the cases of the discrete
derivatives (\ref{derivadadiscretap})--(\ref{derivadadiscretas}), as well as
that of a $q$--derivative. In this way we will get different
representations of the same Lie algebra. We conclude with  some
remarks and comments.

%%%%%%%%%%%%%%%%%%%%%%%%%%%%% SECTION 2 %%%%%%%%%%%%%%%%%%%%%%%%%%%%%%%%
\sect{Discrete heat equation}\label{calor}
%%%%%%%%%%%%%%%%%%%%%%%%%%%%%%%%%%%%%%%%%%%%%%%%%%%%%%%%%%%%%%%%%%%%%%%%

Let us consider the  second order difference equation
$$
(\Delta_{t}-\Delta_{xx})u(x)=0 ,
$$
as a discretization of the heat equation. Since it is linear, we 
can consider an evolutionary vector field of the form
\be\label{evolvector}
X_e\equiv Q\partial_u=(\tau\Delta_{t} +\xi \Delta_{x} u+ f u) \partial_u ,
\ee
where $\tau$, $\xi$ and $f$ are (operator valued) functions of $x, t, T_x$
$T_t$, $\sx$  and $\st$. The determining equation is
$$
\Delta_{t}^T Q -  \Delta^T_{xx} Q|_{\Delta_{xx}u=\Delta_{t} u}=0 ,
$$
whose explicit expression is
\be\label{ecuacioncalordet1}
\Delta_{t}(\xi\Delta_{x}u) +\Delta_{t}(\tau\Delta_{t}u)+ \Delta_{t}(fu)
-[\Delta_{xx}(\xi\Delta_{x}u) +\Delta_{xx}(\tau\Delta_{t}u)+ \Delta_{xx}(fu)] 
|_{\Delta_{xx}u=\Delta_{t} u}=0 . 
\ee

Only when  expression (\ref{ecuacioncalordet1}) is developed one  needs to apply
a Leibniz rule and, hence, the results will depend from the definition of
the corresponding discrete derivative. 
We propose a Leibniz rule having the form
\be\label{Leibnizgeneral}
\Delta_x\left(f(x)g(x)\right)= f(x) \Delta_x g(x)+
D_x(f(x))g(x),
\ee
where $D_x(f(x)) = [\Delta_x,f(x)]$ is a function of $x, T_x$ and $\sx$
(similarly for $D_t(f(t))$).

Using the general rule (\ref{Leibnizgeneral}) for an arbitrary discrete
derivative we obtain from (\ref{ecuacioncalordet1}), equating to zero  the
coefficients of $\Delta _{xt} u$, $\Delta _{t} u$,  $\Delta _{x} u$ and $u$,
respectively, the following set of  determining equations
\be\begin{array}{l}\label{ecuacacionesdeterminantescalorgeneral}
D_{x}(\tau)=0 ,\\[0.2cm]
D_{t}(\tau) -2 D_{x}(\xi)=0,\\[0.2cm]
D_{t}(\xi) - D_{xx}(\xi)  -2 D_{x}(f)=0 ,\\[0.2cm]
D_{t} (f)  -D_{xx}(f)=0 
\end{array}\ee 
where $D_{xx}(f)=D_{x}(D_{x}(f))$.
Next, starting from (\ref{ecuacacionesdeterminantescalorgeneral}) we
will study separately the cases for $\Delta^\pm$ and
$\Delta^s$.

%%%%%%%%%%%%%%%%%%%%%%%%%%%%
\subsect{Symmetries for right (left) discrete derivatives}
\label{simetriasdiscetasderecha}
%%%%%%%%%%%%%%%%%%%%%%%%%%%%

Choosing as in Ref. \cite{decio,javier} the derivative $\Delta^+$ and, 
consequently, the Leibniz rule 
$$
\Delta^+ (fg)= f\Delta^+ g + \Delta^+ (f) Tg ,
$$
we get from (\ref{ecuacacionesdeterminantescalorgeneral}), 
\be\begin{array}{l}\label{ecuacacionesdeterminantescalor}
\Delta_{x}^+\tau=0 ,\\[0.2cm]
(\Delta_{t}^+\tau)T_t -2 (\Delta^+_{x}\xi) T_x =0 ,\\[0.2cm]
(\Delta_{t}^+\xi) T_t - (\Delta^+_{xx}\xi) T_x^2  -2 (\Delta^+_{x}f) T_x =0 ,
\\[0.2cm]
(\Delta^+_{t} f) T_t -(\Delta^+_{xx}f)T_x^2=0 .
\end{array}\ee 
The solution of  (\ref{ecuacacionesdeterminantescalor}) gives
\be\label{calorsolucionesdeterminingequation}
\begin{array}{l}
\tau=t^{(2)}\tau_2 + t \tau_1 +\tau_0 ,\\[0.2cm]
\xi=\frac 12 x(\tau_1+2t\tau_2)T_tT_x^{-1}+t\xi_1+\xi_0, \\[0.2cm]
f=\frac 14 x^{(2)} \tau_2T_t^2T_x^{-2}+\frac 12 t\tau_2T_t+
\frac 12 x\xi_1T_tT_x^{-1}+\gamma ,
\end{array}\ee 
where $\tau_0,\  \tau_1,\  \tau_2,\  \xi_0,\ \xi_1$ and $\gamma$ are 
arbitrary functions of $T_x$, $T_t$, and the spacings $\sigma_x$ and
$\sigma_t$. The notation $x^{(n)}$, $t^{(n)}$  is for Pochhammer symbols;
for instance
$$
x^{(n)}=x(x-\sigma_x)\dots (x-(n-1)\sigma_x).
$$
By a suitable choice of the functions  $\tau_i,\ \xi_i,$ and $\gamma$ we get 
the following symmetries \cite{dno00}
\be\label{caloralgebra}
\begin{array}{l}
P_0= (\Delta_t u)\partial_u ,\\[0.2cm]
P_1=(\Delta_x u)\partial_u ,\\[0.2cm]
W= u\partial_u , \\[0.2cm]
B=(2tT_t^{-1}\Delta_x u +xT_x^{-1} u )\partial_u ,\\[0.2cm] 
D= (2tT_t^{-1}\Delta_t u +xT_x^{-1}\Delta_x u + \frac 12 u)\partial_u ,
\\[0.2cm]  
 K=(t^2T_t^{-2}\Delta_t u - \sigma_t t T_t^{-2}\Delta_t u +
txT_t^{-1}T_x^{-1}\Delta_x u  \\[0.1cm] 
\qquad \qquad +  \frac 14 x^2 T_x^{-2} u  - \frac14 \sigma_x x T_x^{-2}u
+\frac12 tT_t^{-1} u)\partial_u .
\end{array}\ee 
Let us note that the above discrete symmetries have a well defined limit
when $\sigma_x,\sigma_t\to 0$, which  leads to the symmetries of the
continuous heat equation. Also, it can be checked that with this choice, the
symmetries (\ref{caloralgebra}) close  a 6--dimensional Lie algebra
isomorphic to the  symmetry algebra of the continuous heat equation valid for
any value of $\sigma_x,\sigma_t$.
\bigskip

A second choice for the discrete derivative is $\Delta^-$. The Leibniz 
rule becomes
\be\label{Leibnizrulem}
\Delta^- (fg)= f\Delta^- g + \Delta^- (f) T^{-1}g .
\ee 
It gives the same results (\ref{calorsolucionesdeterminingequation}) and
(\ref{caloralgebra})  provided 
we make the substitution
$T \rightarrow T^{-1}$.

%%%%%%%%%%%%%%%%%%%%%%%%%%%%%%
\subsect{Symmetries for symmetric discrete derivatives}
\label{simetriasdiscetassimetrica}
%%%%%%%%%%%%%%%%%%%%%%%%%%%%%%

Next, let us consider the case of the symmetric derivative $\Delta^s$
(\ref{derivadadiscretas}). The commutator
$$
[\Delta^s_x,x]=\frac{T_x+T_x^{-1}}{2}
$$
 can always rewritten by introducing a function
$\bsx=\b^s(T_x)=2(T_x+T_x^{-1})^{-1}$ as
$$
[\Delta_x,x\bx]=1 .
$$
This fact will help us in the computation of general commutators, so
\be\label{commutadorfs}
[\Delta^s_x,f(x)\bsx]=\left(\Delta^s_xf(x)\right)T_x\bsx
+\left(T_x^{-1} f(x)-f(x)\right)\bsx \Delta^s_x.
\ee
From  relation (\ref{commutadorfs}) we get the explicit Leibniz rule
$$
\Delta^s_x\left(f(x)g(x)\right)= f(x) \Delta^s_x g(x) + \left[\right. 
\frac{1}{\sigma_x}\left((T_x^{-1}-1)f(x)\right)
\left(T_x-(\bsx)^{-1}\right)
  +\left. \left(\Delta^s_xf(x)\right)T_x\right]g(x).
$$
This formula  allows us to  write explicitly the 
determining equations
(\ref{ecuacacionesdeterminantescalorgeneral}). Their  solution is given by
\cite{dno00}
$$\label{ecuacacionesdeterminantescalor4soluciones}\begin{array}{l}
\tau^s = t^{(2)}\tau_2 +t\tau_1+\tau_0 ,
\\[0.2cm]
\xi^s = \frac {1}{2}x \left(2t\tau_2 + \tau_1 + \st T_t^{-1} \bst \tau_2\right)
(\bst)^{-1}\bsx +t\xi_1+\xi_0,
\\[0.2cm]
f^s = \frac{1}{4} x^{(2)} \tau_2(\bsx)^2 (\bst)^{-2}  
{+} \frac{1}{2}x\xi_1 \bsx (\bst)^{-1} 
{+} \frac 14 x\sx\tau_2 T_x^{-1} (\bsx)^3(\bst)^{-2} {+} \frac12 t \tau_2
 (\bst)^{-1}  {+} f_0,
\end{array}$$
where $\tau_2$, $\tau_1$, $\tau_0$, $\xi_1$, $\xi_0$ and $f_0$ are arbitrary 
functions of $T_x,\ T_t,\ \sx$ and $\st$.
Now, from  these solutions and (\ref{evolvector})  we obtain,
with a suitable choice of  $\tau_2$, $\tau_1$, $\tau_0$, $\xi_1$, $\xi_0$ and
$f_0$, the following symmetries
\be\label{caloralgebrasimetrica}
\begin{array}{l}
P_0^s= (\Delta^s_t u)\partial_u ,\\[0.2cm]
P_1^s=(\Delta^s_x u)\partial_u , \\[0.2cm]
W^s= u\partial_u , \\[0.2cm]
B^s=(2t\bst\Delta^s_x u + x\bsx u )\partial_u ,  \\[0.2cm]
D^s= (2t\bst\Delta^s_t u +x\bsx\Delta^s_x u + \frac 12 u)\partial_u ,
\\[0.2cm]
K^s=((t^2(\bst)^{2}- t{\sigma_t}^2(\bst)^3{\Delta^s_t})\Delta^s_t u  +
tx\bst\bsx\Delta^s_x u   \\[0.1cm]
\qquad \quad  
- \frac 14 x{\sigma_x}^2(\bsx)^3\Delta_x^s u + \frac 14 x^2 (\bsx)^{2} u
+ \frac 12 t\bst u)\partial_u .
\end{array}\ee 
These symmetries close the same  6--dimensional Lie algebra generated by the
operators  (\ref{caloralgebra}), and have a well defined continuous limit.

%%%%%%%%%%%%%%%%%%%%%%%%%%%%
\subsect{Symmetries for $q$--derivatives}
\label{qsimetriasdiscetas}
%%%%%%%%%%%%%%%%%%%%%%%%%%%%

In the following we shall extend the preceding method to the case of
$q$--derivatives and $q$--symmetries.  We deal  briefly with a $q$--discretized
heat equation where the
$q$--difference operator is defined by
$$
\Delta^q_x =\frac{1}{(q_x-1)x}(T_x-1) ,
$$
with the help of a $q$--shift operator 
$$
T_x =e^{q_x x\partial_x},\qquad  T_x f(x) = f(q_x x).
$$
In this case  we have the commutator 
$$
[\Delta^q_x,x]= T_x.
$$
If we look for a function $\bx(T_x)$ satisfying 
$$
[\Delta^q_x,\bx x] =1,
$$
a formal solution is
$$
\bx(T_x) = (q_x-1)\frac{x\partial_x}{T_x - 1}.
$$
Thus, we can perform a change of basic operators
$$
\{x,T_x\} \to \{\tilde x,\Delta^q_x\},\qquad \tilde x= \bx(T_x) x,
$$
so that, formally, we can express any function $f(x,T_x)$ as
$$
f(x,T_x) = \widetilde f(\tilde x, \Delta^q_x).
$$
In this way the determining equations
(\ref{ecuacacionesdeterminantescalorgeneral}) take the form
\begin{eqnarray*}
\widetilde \tau_{\tilde x}=0 ,\\[0.2cm]
\widetilde \tau_{\tilde t} -2 \widetilde \xi_{\tilde x}=0,\\[0.2cm]
\widetilde \xi_{\tilde t} - \widetilde \xi_{\tilde x \tilde x}  -2
\widetilde f_{\tilde x}=0 ,\\[0.2cm] \widetilde f_{\tilde t}  -
\widetilde f_{\tilde x \tilde x}=0 .
\end{eqnarray*}
Therefore, we can give a set of solutions that have a similar appearance to the
classical symmetries
\begin{eqnarray*}
&&P_0^q= (\Delta^q_t \,u)\partial_u ,\\[0.1cm]
&&P_1^q=(\Delta^q_x \,u)\partial_u , \\[0.1cm]
&&W^q= u\partial_u , \\[0.1cm]
&&B^q=(2\bt t\Delta^q_x \,u + \bx x \,u )\partial_u ,  \\[0.1cm]
&&D^q= (2\bt t\Delta^q_t\, u + \bx x\Delta^q_x \,u + \frac 12 \,u)\partial_u ,
\\[0.1cm]
&&K^q = (\gamma_t t^2 \Delta^q_t \,u +\bx\bt t x \Delta^q_x \,u +
\frac14 \gamma_x x^2 \,u + \frac 12 \bt t\, u )\partial_u,
\end{eqnarray*}
where  $\gamma_x =
\left[\frac{(q-1)x\partial_x}{T_x-1}\right]
\left[\frac{(q-1)(-1+x\partial_x)}{q_x^{-1}T_x-1}\right]$ (for $\gamma_t$
replace $x$ by $t$) .

%%%%%%%%%%%%%%%%%%%%%%%%%%%%% SECTION 5 %%%%%%%%%%%%%%%%%%%%%%%%%%%%%%%
\sect{Conclusions}\label{conclusiones}
%%%%%%%%%%%%%%%%%%%%%%%%%%%%%%%%%%%%%%%%%%%%%%%%%%%%%%%%%%%%%%%%%%%%%%%

We remark that the key point in obtaining the explicit determining equations
(\ref{ecuacacionesdeterminantescalorgeneral}) and, consequently, the discrete
symmetries is the use of the Leibniz rule defined in (\ref{Leibnizgeneral}). Of
course, this approach is not the only possibility; in fact, it must be checked
whether it works correctly or not for each case under study. The choice of a
commutator in order to define the Leibniz rule implicitly leads us to Lie
symmetries, since the natural algebraic structure will be given also in terms of
commutators.

Some of the above result deserves some comments. The symmetries associated
to the  symmetric derivatives (Section \ref{simetriasdiscetassimetrica}) include
functions $\bst$ ($\bsx$) of $T_t$ ($T_x$) that can only be understood as infinite
series expansions. Therefore, not all the symmetries
(\ref{caloralgebrasimetrica}) have a local character, in the sense that they
are not (finite) polynomials in the operators $T_t^{\pm1},\ T_x^{\pm1}$.
Note that although these discrete symmetries  give rise to the
classical symmetries in the limit
$\sigma_x\to 0,\sigma_t\to 0$, one of them, $K^s$,
also includes surprisingly a term in $(\Delta_t)^2$, which vanishes in the
continuous limit since it contains a factor $\sigma_t^2$.

Something similar happens  with the $q$--symmetries (Section
\ref{qsimetriasdiscetas}): they have  also a highly non-local
character. The origin of this unpleasant feature is that the basic commutator
 $[\Delta^q_x,\bx x] =1$ needs a non-local function $\bx(T_x)$. If
we want to investigate local symmetries it is necessary to use another commutator 
free of  this problem. For instance, we could  take  as starting point the
$q$--commutator: $[\Delta^q_x, x T_x^{-1}]_{q_x} =1$, where $[A,B]_{q_x} = AB-
q_x^{-1} BA$. However, this treatment would lead us to a $q$--algebra \cite{vinet}
which is out of our present scope.

%%%%%%%%%%%%%%%%%%%%%%%%%%%%%%%%%%%%%%%%%%%%%%%%%

Let us insist that the procedure here exposed can be
straightforwardly applied to other discretizations such as the wave equation
\cite{nn96} or even equations including a potential term as long as we keep
inside the field of linear equations. Non-linear equations needs 
additional improvements in order to have reasonable determining equations.

%%%%%%%%%%%%%%%%%%%%%%%% ACKNOWWLEDGMENTS %%%%%%%%%%%%%%%%%%%%%%%%%%%%%
\section*{Acknowledgments}
%%%%%%%%%%%%%%\%%
This work has been partially supported by 
DGES of the  Ministerio de Educaci\'on y Cultura of Spain under 
Projects PB98-0360 and the Junta de Castilla y Le\'on (Spain).  

%%%%%%%%%%%%%%%%%%%%%% BIBLIOGRAPHY %%%%%%%%%%%%%%%%%%%%%%%%%%%

%%%%%%%%%%%%%%%%%%%%
\end{document}